\documentclass[a4paper]{article}
\usepackage[utf8]{inputenc}
\usepackage[T1]{fontenc}
\usepackage[numbers,compress,merge]{natbib}
\usepackage[colorlinks=true]{hyperref}
\usepackage{graphicx}
\usepackage{amsmath,amssymb}
\usepackage{subfigure}
\usepackage{color,xcolor}
\usepackage[varg]{txfonts}
\usepackage[affil-it]{authblk}
\newcommand{\Ai}{A^{\lambda,\text{in}}_{\omega l}}

\newcommand{\Ar}{A^{\lambda,\text{ref}}_{\omega l}}

\begin{document}
 
\title{Electromagnetic absorption, emission and scattering spectra of black holes with tidal charge}

\author{Ednilton S. de Oliveira\footnote{ednilton@ufpa.br}}
\affil{Faculdade de F\'isica, Universidade Federal do Par\'a,
	66075-110, Bel\'em, Par\'a, Brazil}

\date{\today}

\maketitle

\begin{abstract}
The effects of the tidal charge of black holes predicted in the Randall-Sundrum brane-world scenario on the electromagnetic cross sections and emission rate is studied. Although quantitatively different, the cross sections and emission rates of such black holes are similar to the ones of Schwarzschild black holes. The increase of the tidal-charge intensity makes the absorption cross sections increase and the interference-fringe widths of the differential scattering cross sections narrow. The electromagnetic emission rate increases initially with the increase of the tidal-charge intensity, but reaches a maximum. This results from the combination of opposite behaviors of the absorption cross section and the black hole temperature, as this last decreases with the increase of the tidal-charge intensity. It is shown that the cross sections obtained via the partial-wave method agree well with the high-frequency approximations, and can also been simulated by their scalar analogues in certain limits.
\end{abstract}


\section{Introduction}

The first image of a black hole shadow has recently been obtained by the Event Horizon Telescope~\cite{eht2019aj_l1,*eht2019aj_l2,*eht2019aj_l3,*eht2019aj_l4,*eht2019aj_l5,*eht2019aj_l6}. This accomplishment, together with the gravitational-wave detections of black hole binary mergers~\cite{ligo2016prl116_061102,*ligo2016prl116_241103,*ligo2017prl118_221101}, is a clear evidence of the existence of black holes and how important such objects are to understand our Universe. These observations have also in common the fact that they rely on theoretical simulations to interpret physically the produced data. Indeed, long before the detections, a vast number of theoretical simulations had given a preliminary sample of what astronomers should observe in the galaxy centers~\cite{Broderick2006aj636_l109,Moscibrodzka2009aj706_497,Dexter2012mnras421_1517,Dibi2012mnras426_1928,Chan2015aj799_1,Moscibrodzka2016aa586_a38,Porth2016cac4_1,Chael2018mnras,Ryan2018aj864_126} and from binary black hole mergers~\cite{Buonanno2000prd62_064015,Centrella2010rmp82_3069,Taracchini2012prd86_024011,Pan2011prd84_124052,Littenberg2013prd87_104003,Taracchini2014prd89_061502,Purrer2014cqg31_195010}.

Considering the importance of black holes and the increasing attention such systems have been receiving, studying every aspect of them is of great importance. Among these aspects, we have to consider black holes described in alternatives to General Relativity, once such objects can help to constrain the set of models which best describe gravitation. Here we study the properties of one of these alternative systems: black holes with tidal charge~\cite{Dadhich2000plb487_1}, predicted in the context of the Randall--Sundrum brane-world scenario~\cite{Randall1999prl83_3370,Randall1999prl83_4690}.

Let us briefly remark the main points about black holes with tidal charge. They are described by the following geometry\footnote{Here we use $c = G = \hbar = 1$.}:
\begin{equation}
 ds^2 = f(r)dt^2-f(r)^{-1}dr^2 - r^2(d\theta^2+\sin^2\theta d\phi^2),
\label{metric}
\end{equation}
with
\begin{equation}
 f(r)= 1 - \frac{2M}{r} + \frac{\beta}{r^2},
\end{equation}
where $M$ is the black hole mass and $\beta \equiv q M^2 < 0$  represents a consequence of the fifth dimension on the 3-brane, populated by Standard--Model fields, and it has been called of ``tidal charge''. We remark the fact that metric~\eqref{metric} coincides with the one for Reissner-Nordström black holes~\cite{Chandra1983} if $0 < q \le 1$.

Studying black holes presented in Eq.~\eqref{metric} is important because it help us to understand some of the consequences of the Randall--Sundrum models. Particularly their first model~\cite{Randall1999prl83_3370}, usually referred to as RS1, is very important once it gives one of the possible solutions to the hierarchy problem. Indeed, searches for microscopic black holes at the LHC have been carried out considering the RS1 model~\cite{ATLAS2016jhep03_041,ATLAS2016plb754_302,CMS2017plb774_279,ATLAS2018epjc78_102,cms2018jhep2018_42}, as well as the Arkani-Hamed--Dimopoulos--Dvali model~\cite{Arkani1998plb429_263,Antoniadis1998plb436_257}.

In Ref.~\cite{deOliveira2018epjc78_876} we computed the cross sections of black holes with tidal charge considering the simple model of massless scalar plane waves. There, we have shown that the tidal charge plays important roles in phenomena which take place near the black hole, once it alters the sizes of the event horizon and the photon sphere. As consequences, the cross sections change considerably with the change of $q$. The absorption cross section of scalar particles becomes bigger and the differential scattering cross sections present narrower interference fringes for black holes with more intense tidal charges. Despite this, the overall behavior of the scalar cross sections is similar to other kind of black holes, specially to the ones of Schwarzschild~\cite{Sanchez1976jmp17_688,Sanchez1976prd16_937,Sanchez1978prd18_1030,Sanchez1978prd18_1798} and Reissner-Nordström~\cite{Jung2005npb717_272,Crispino2009prd79_064022} black holes. Numerical results agree well with all analytical approximations and it becomes difficult to distinguish between Schwarzschild, Reissner--Nordström and black holes with tidal charge when considering scattering in the weak-field limit. Other aspects of such black holes, as shadows~\cite{Schee2008ijmpd18_983,Amarilla2011pprd85_064019,Abdujabbarov2017prd96_084017,Eiroa2017epjc78_91}, quasinormal modes~\cite{Toshmatov2016prd93_124017}, and their effect on particle dynamics~\cite{Abdujabbarov2010prd81_044022,Rahimov2011ass335_499,Shaymatov2013prd88_024016} have been considered by different authors.

On the present work we extend the previous analysis made in Ref.~\cite{deOliveira2018epjc78_876} by contemplating now the electromagnetic case. We compute numerically the cross sections of black holes with tidal charge considering circularly polarized electromagnetic waves. We compare our results with analytical results, as geodesic and glory scattering, in order to test their consistency. Comparisons of the main results with the ones for Schwarzschild black holes are presented in order to highlight the effects of the tidal charge. We also compare results for both electromagnetic and scalar cases showing that the scalar case can mimic the results of the electromagnetic waves in some limits.

This paper is divided as follows: In Sec.~\ref{sec:waves} we describe the overall behavior of the electromagnetic waves around black holes with tidal charge; in Sec.~\ref{sec:results} we show our results and compare them with the equivalent results for the Schwarzschild black holes, as well as their high-frequency and scalar equivalents. Although we present cross sections based on the geodesic and glory analysis, we indicate Ref.~\cite{deOliveira2018epjc78_876} to the reader who wishes for a detailed analysis of such approximations. In Sec.~\ref{sec:conclusion} we present our final remarks and conclusions.

\section{Electromagnetic-wave scattering}
\label{sec:waves}

Here we make a succinct description of electromagnetic-wave propagation around black holes with tidal charge, once such analysis is very similar to the Schwarzschild case, which was done, for example, in Ref.~\cite{Crispino2007prd75_104012}. Following this reference, the electromagnetic field can be described in the modified Feynman gauge via the Lagrangian density
\begin{equation}
\mathcal{L} = \sqrt{-g} \left[-\frac{1}{4} F_{\mu\nu}F^{\mu\nu} - \frac{1}{2} H^2\right],
\label{lag}
\end{equation}
where $g = \det(g_{\mu\nu})$, $H \equiv \nabla^\mu A_\mu + K^\mu A_\mu$, with $K^\mu = (0, f'(r), 0, 0)$. In this gauge, the physical modes can be written as: 
\begin{eqnarray}
A^{(I\omega lm)}_{\mu} = &(0, \frac{\varphi^{I}_{\omega l}(r)}{r^{2}}
Y_{lm}, \frac{f}{l(l+1)} \frac{d}{dr}[\varphi^{I}_{\omega l}(r)]
\partial_{\theta}Y_{lm},
\frac{f}{l(l+1)}
\frac{d}{dr}[\varphi^{I}_{\omega l}(r)]
\partial_{\phi}Y_{lm} ) e^{-i\omega t},
\label{AI}
\end{eqnarray}
\begin{equation}
A^{(II\omega lm)}_{\mu} = (0, 0, \varphi^{II}_{\omega l}(r)
Y_{\theta}^{(lm)}, \varphi^{II}_{\omega l} (r) Y_{\phi}^{(lm)} )
e^{-i\omega t},
\label{AII}
\end{equation}
where $Y_{lm}$ are the scalar spherical harmonics and $Y_a^{(lm)}$ are the vector spherical harmonics~\cite{Higuchi1987cqg4_721}. The radial functions $\varphi^{I}_{\omega l}$ and $\varphi^{II}_{\omega l}$ obey the same radial equation in the case of black holes with tidal charge~\cite{Toshmatov2016prd93_124017,Molina2016prd93_124068}, similarly to what happens around Schwarzschild black holes~\cite{Crispino2007prd75_104012,Crispino2009prl102321103}. Therefore, we can synthesize the description of the radial functions as $\varphi^{\lambda}_{\omega l}$ ($\lambda = I,II$) by writing the following radial equation:
\begin{equation}
f\frac{d}{dr} \left[f \frac{d}{dr}\varphi^{\lambda}_{\omega l}(r)\right] + \left[\omega^{2} - V(r) \right]\varphi^{\lambda}_{\omega l}(r)=0,
\label{rad_eq}
\end{equation}
where the effective potential reads:
\begin{equation}
V(r) = f\frac{l(l+1)}{r^{2}}.
\label{pot}
\end{equation}
It can be straightforwardly verified that $V(r) \to 0$ for $r \to r_h$, where $r_h$ is the event horizon radius, and for $r \to \infty$. From this we can write the asymptotic behaviors of $\varphi^\lambda_{\omega l}$, which are better expressed in the terms of the tortoise coordinate, defined as $d/dr_* \equiv f\,d/dr$:
\begin{equation}
\varphi^{\lambda}_{\omega l}(r_*)\approx \left\{
\begin{array}{lr}
A^{\lambda,\text{tr}}_{\omega l}
e^{-i\omega r_*} & (r_* \rightarrow -\infty),\\
A^{\lambda,\text{in}}_{\omega l} e^{-i\omega r_*} + A^{\lambda,\text{ref}}_{\omega l}e^{i\omega r_*} & (r_* \rightarrow +\infty).
\end{array}
\right.
\label{asymp}
\end{equation}
The superscript labels ``tr'', ``in'' and ``ref'' stand for transmitted, incident and reflected parts of the scattered wave, respectively, so that the reflection and transmission coefficients can be respectively defined as:
\begin{equation}
 \mathcal{R}^{\lambda}_{\omega l} \equiv \left|\frac{A^{\lambda,\text{ref}}_{\omega l}}{A^{\lambda,\text{in}}_{\omega l}}\right|^2,
 \label{R}
\end{equation}
and
\begin{equation}
 \mathcal{T}^{\lambda}_{\omega l} \equiv \left|\frac{A^{\lambda,\text{tr}}_{\omega l}}{A^{\lambda,\text{in}}_{\omega l}}\right|^2.
 \label{T}
\end{equation}
These coefficients rule the amount of reflected and absorbed parts of the scattered modes and, therefore, are directly related to the absorption cross section, the evaporation rate and to each other via $\mathcal{R}^{\lambda}_{\omega l} + \mathcal{T}^{\lambda}_{\omega l} = 1$. From Eq.~\eqref{asymp} we can also define the phase shifts, which play a key role in the differential scattering cross sections:
\begin{equation}
 e^{2i\delta^\lambda_l(\omega)} = (-1)^{l+1} \frac{A^{\lambda,\text{ref}}_{\omega l}}{A^{\lambda,\text{in}}_{\omega l}}.
 \label{ps}
\end{equation}

The electromagnetic absorption cross section of spherically symmetric black holes can be expressed generally as~\cite{Crispino2007prd75_104012,Crispino2009prd80_104026}:
\begin{equation}
\sigma_\text{abs} = \sum\limits_{l=1}^{\infty}{\sigma_\text{abs}^{(l)}} ,
\label{ACS}
\end{equation}
which is composed by the sum of partial absorption cross sections, which are:
\begin{equation}
{\sigma_\text{abs}^{(l)}} =
\frac{\pi}{2\omega^{2}} {
\sum\limits_{\lambda=I,II}\left( 2l+1 \right)\mathcal{T}^{\lambda}_{\omega l}}.
\label{PACS}
\end{equation}
The electromagnetic emission rate in a frequency interval between $\omega$ and $\omega + d\omega$ is given by~\cite{Hawking1975cmp43_199}
\begin{equation}
\frac{dE}{dt} = \frac{\omega^3}{2\pi^2\left(e^{\omega/T}-1\right)} \sigma_\text{abs} d\omega,
\label{hr}
\end{equation}
where $T$ is the black hole temperature, that in the present case is~\cite{Toshmatov2016prd93_124017}
\begin{equation}
T = \frac{\sqrt{M^2-q}}{2\pi(M+\sqrt{M^2-q})^2}.
\label{temp}
\end{equation}

The differential scattering cross section of circularly polarized waves can be expressed as~\cite{Fabbri1975prd12_933}:
\begin{equation}
 \frac{d\sigma_\text{el}}{d\Omega} = \frac{1}{8\omega^2} \left\{ \left|
\sum\limits_{l=1}^{\infty}
\frac{2l+1}{l(l+1)}\left[e^{2i\delta_{l}^{I}(\omega)}
T_l (\theta) + e^{2i\delta_{l}^{II} (\omega)}\pi_l (\theta) \right]
\right|^2 +
\left|\sum\limits_{l=1}^{\infty} \frac{2l+1}{l(l+1)} \left[
e^{2i\delta_{l}^{I} (\omega)} \pi_l (\theta) +
e^{2i\delta_{l}^{II} (\omega)} T_l (\theta) \right] \right |^2
\right\},
\label{scs}
\end{equation}
where
\begin{equation}
 \pi_l(\theta) \equiv \frac{P_l^1(\cos\theta)}{\sin\theta}, \qquad
T_l (\theta) \equiv \frac{d}{d\theta } P_l^1 (\cos\theta),  \label{eq:pi-T}
\end{equation}
being $P_l^1$ associated Legendre functions. The differential scattering cross section can be written alternatively as a sum of a helicity-preserving with a helicity-reversing scattering amplitude~\cite{Crispino2014prd90_064027}. This is interesting when the two kinds of polarized modes are scattered differently, e.g. when there is an electrostatic field filling the space, as it occurs around Reissner-Nordström black holes. One of the effects of a helicity-reversing scattering is a nonzero flux of scattered waves in the backward direction. However, this is not the case here, as well as in the Schwarzschild case~\cite{Mashhoon1973prd7_2807}, once $\delta_l^I = \delta_l^{II}$ because both modes are subjected to the same effective scattering potential.

\section{Results}
\label{sec:results}

The present section contains a selection of results for the electromagnetic absorption and differential scattering cross sections as well as for the electromagnetic emission spectra of black holes with tidal charge. Such results are obtained by evaluating numerically the radial equation~\eqref{rad_eq} from $r \gtrsim r_h$ to $r \gg r_h$, and then matching the numeric solutions with the asymptotic ones to obtain the necessary coefficients. We can improve precision by writing an alternative asymptotic solution in the far region by noticing that $V \sim l(l+1)/r^2$ for $r \gg r_h$, and therefore:
\begin{equation}
\varphi^{\lambda}_{\omega l} (r_*) \approx \omega r_*\left[(-i)^{l+1}\Ai \, h_l^{(2)}(\omega r_*) + i^{l+1} \Ar \, h_l^{(1)}(\omega r_*)\right],
\label{psi_hankel}
\end{equation}
where $h^{(1)}_l $ and $h^{(2)}_l$ are the spherical Hankel functions of the first and second kinds~\cite{Abramowitz_etal1964}, respectively. This improvement is specially important in the computation of the differential scattering cross sections as it requires a high number of terms in the sums of Eq.~\eqref{scs} which include phase shifts obtained numerically. Furthermore, since such sums are poorly convergent, a convergence method of reduced series similar to the one introduced in Ref.~\cite{Yennie1954pr85_500} has to be applied. This generalized method for electromagnetic differential scattering cross section has been applied in Refs.~\cite{Crispino2009prl102321103,Crispino2014prd90_064027}.

\begin{figure}[!htpb]
\centering
\includegraphics[width=0.5\textwidth]{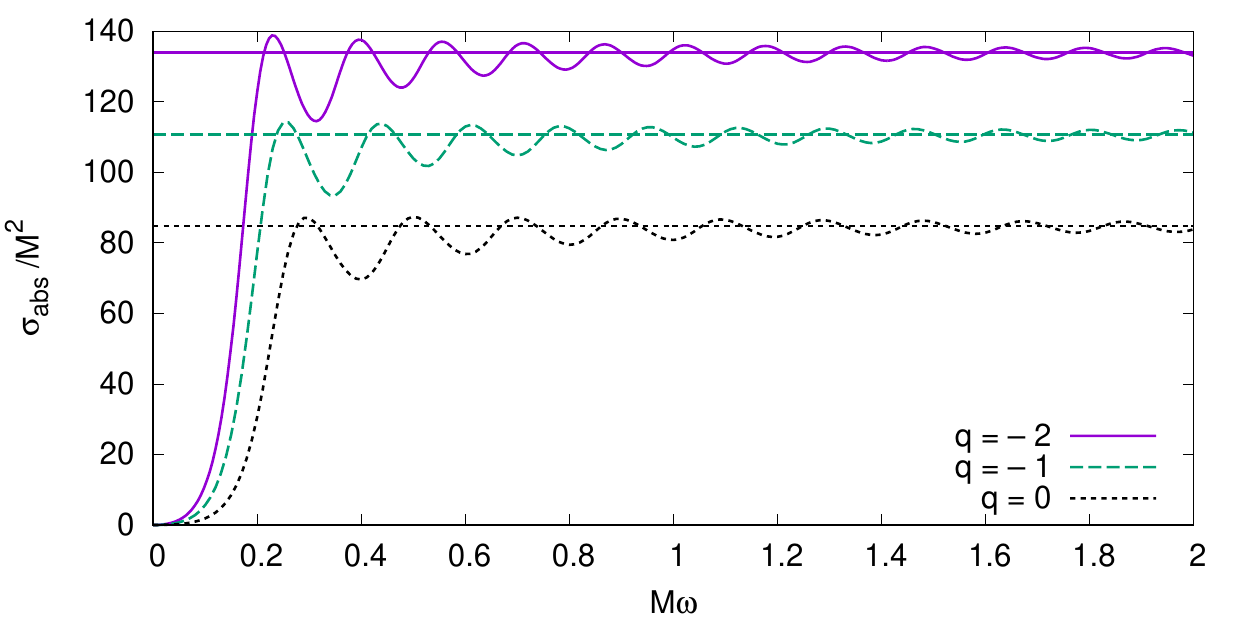}
\caption{Total electromagnetic absorption cross sections of black holes with tidal charge $q = -2,-1$ and Schwarzschild black holes. The straight lines represent the corresponding geometrical-optics limit.}
\label{fig:tacs}
\end{figure}

Figure~\ref{fig:tacs} shows the total electromagnetic absorption cross sections for black holes with $q = -2,-1,0$. As we can see, absorption of black holes with tidal charge is enhanced when compared with the absorption of Schwarzschild black holes ($q = 0$). In the same figure, the straight lines represent the corresponding capture cross section computed via geodesic analysis. It is clear that these results should be recovered in the limit $M\omega \to \infty$, once the partial-wave results present waning oscillations around them as $M\omega$ increases. A similar behavior has been observed in the case of the massless scalar field~\cite{deOliveira2018epjc78_876,Toshmatov2016prd93_124017}.

\begin{figure}[!h]
 \centering
 \includegraphics[width=0.5\textwidth]{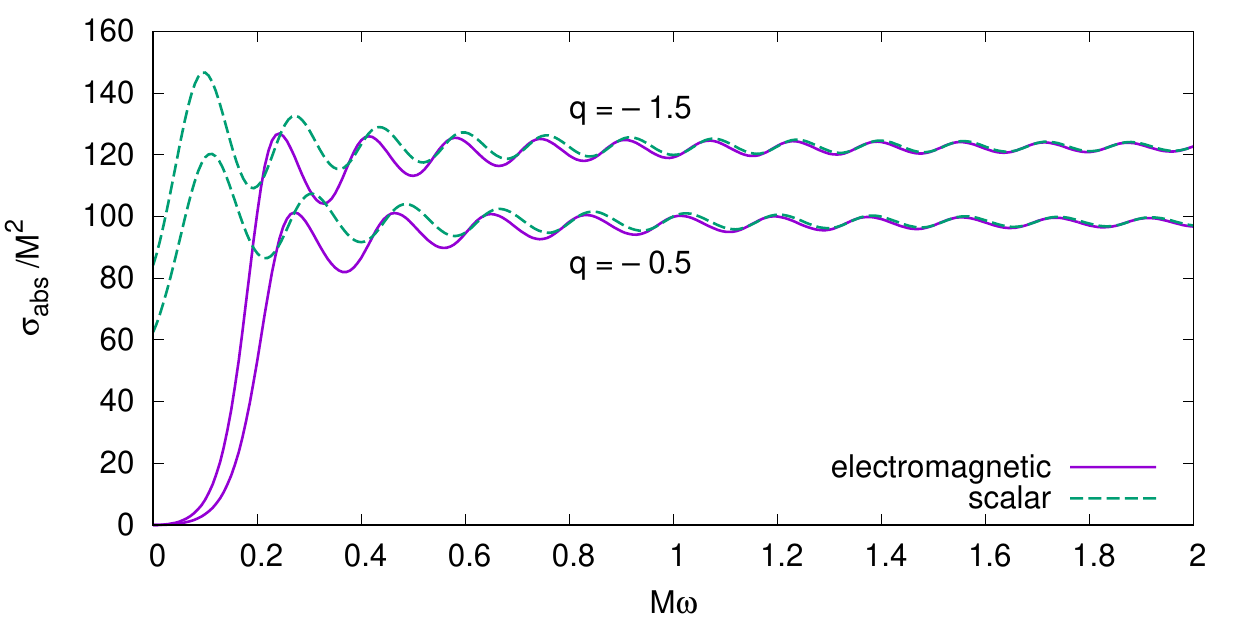}
 \caption{Absorption cross section of black holes with tidal charges $q = -1.5, -0.5$ for spin 0 and 1 massless particles. Agreement between the two cases does not depend on the black hole charge and happens already for intermediate values of frequency, $M\omega \gtrsim 1$.}
 \label{fig:abs_comp}
\end{figure}

Figure~\ref{fig:abs_comp} shows the comparison of the electromagnetic and scalar absorption cross sections of black holes with tidal charges $q = -0.5,-1.5$. We see that the differences are seen mainly in the low-frequency regime, while for high frequencies the results tend to coincide independently of the black hole charge value. In Ref.~\cite{deOliveira2018epjc78_876}, scalar absorption cross section has been shown to excellently agree with the ``sinc approximation''~\cite{Toshmatov2016prd93_124017} for the same values of $q$ used in Fig.~\ref{fig:abs_comp}. Therefore, the results for the electromagnetic case presented here are consistent with this approximation as well.


\begin{figure}[!htb]
\centering
\includegraphics[width=0.5\textwidth]{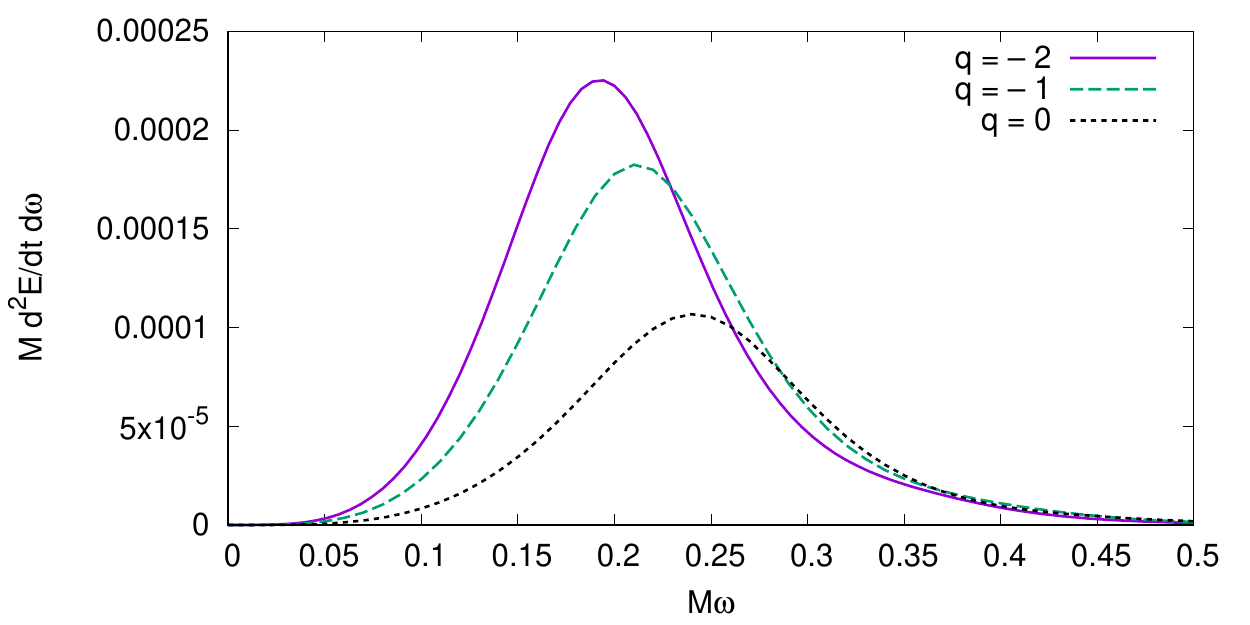}
 \caption{Electromagnetic emission spectra of black holes with tidal charges $q= -2,-1$ and Schwarzschild black holes.}
 \label{fig:hr}
\end{figure}

\begin{figure}[!htb]
 \centering
 \includegraphics[width=0.5\textwidth]{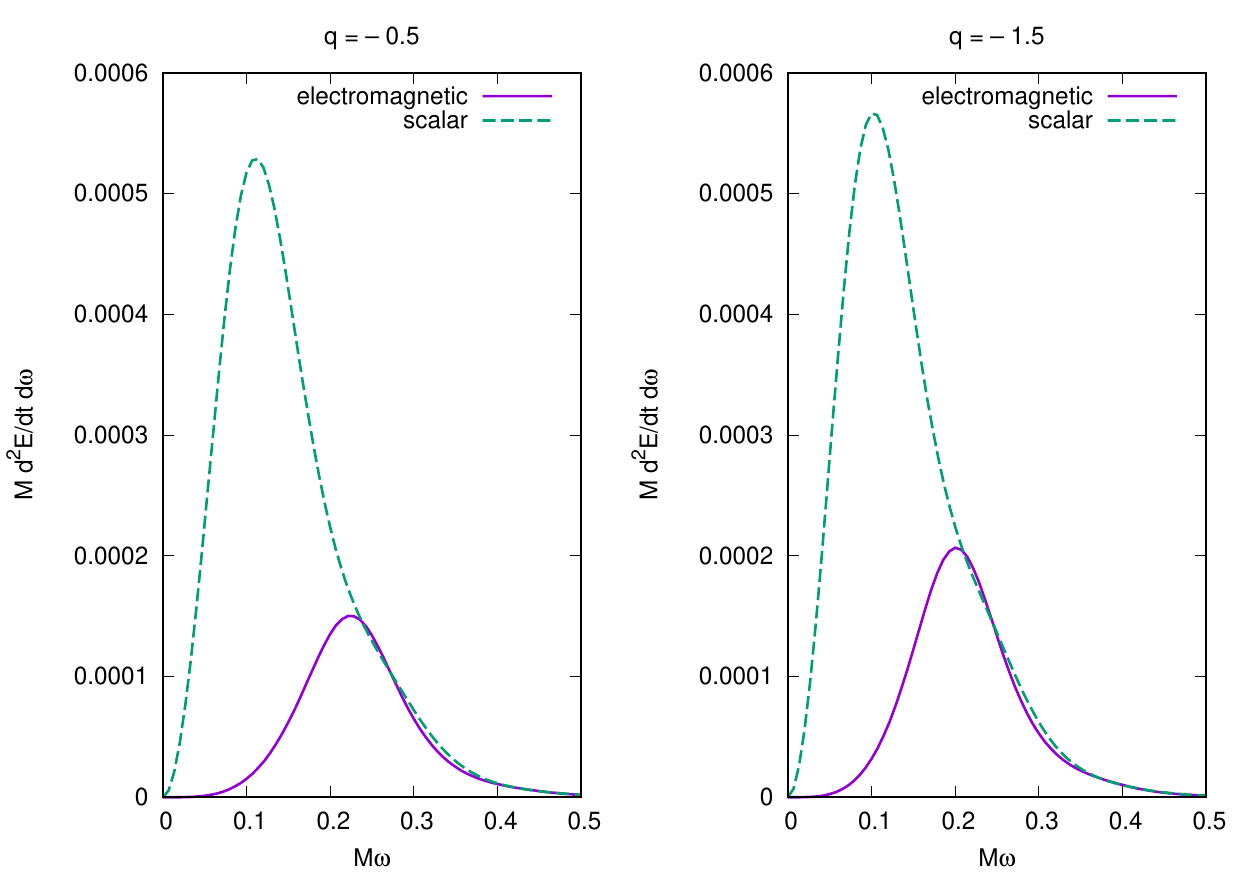}
 \caption{Comparison between the electromagnetic and scalar emission spectra of black holes with tidal charge. As example, we have chosen the cases $q = -0.5$ (left) and $q=-1.5$ (right).}
 \label{fig:comp_hr}
\end{figure}

In Fig.~\ref{fig:hr} we show the emission rate of photons from black holes with different tidal charges, including the Schwarzschild case. The electromagnetic emissions presented in this figure are higher for higher tidal-charge intensities, although the increase of $-q$ reduces the black hole temperature. We should emphasize, however, that comparing the emission of black holes with same mass and more intense tidal charges means comparing black holes with bigger event horizons. The combination of an increasing absorption cross section with a decreasing temperature makes the emission spectra narrows, what indicates that the total emission should ceases to increase if we consider higher tidal-charge intensities. In fact, we have found that the electromagnetic emission rate, considering all frequencies, reaches a maximum value around $q = -4.6$. In Fig.~\ref{fig:comp_hr} we show the comparison of the emission rates of black holes with tidal charges considering both spin 0 and 1 massless particles. Electromagnetic emission is much lower than scalar; with an estimated ratio of 28\% in the $q = -0.5$ case and 34\% for the $q = -1.5$ case.

\begin{figure}[!htpb]
\centering
\includegraphics[width=0.5\textwidth]{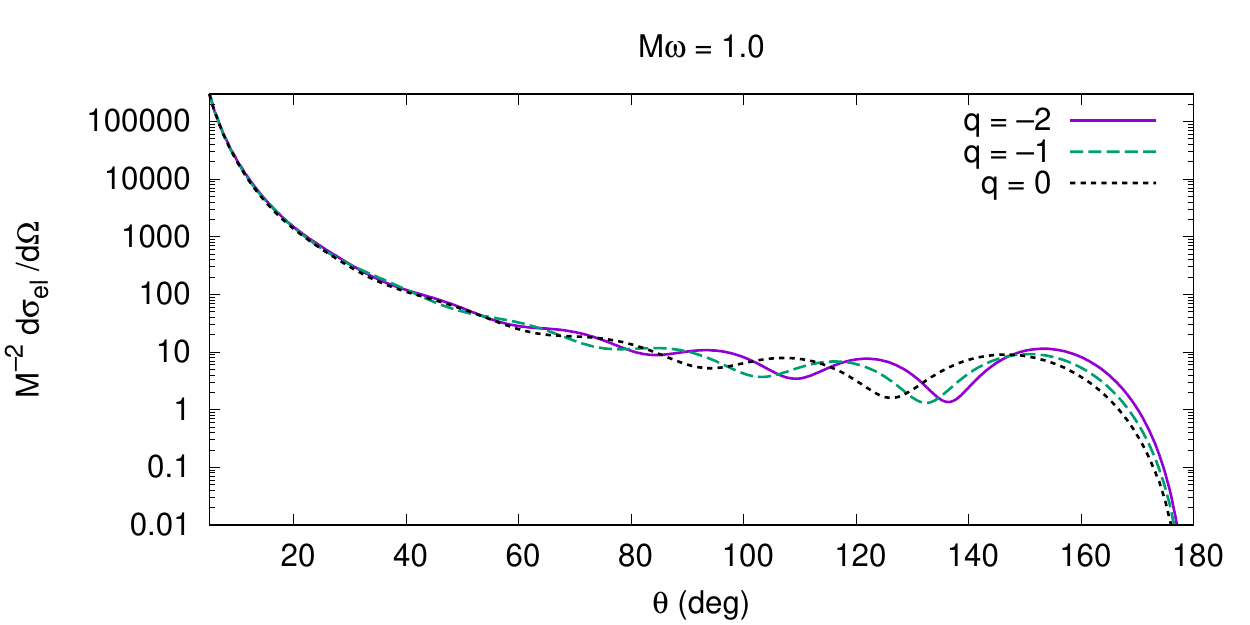}
\caption{Electromagnetic differential scattering cross section for black holes with tidal charges $q = -1,-2$ and Schwarzschild black holes. Here we compare the cases of $M\omega  = 1$.}
\label{fig:scs_w1}
\end{figure}

In Fig.~\ref{fig:scs_w1} we compare the electromagnetic differential scattering cross section for black holes with tidal charges $q = 0,-1,-2$ considering $M\omega = 1.0$. There we see that the tidal charge has visible consequences in the widths of the interference fringes but low influence on the scattered flux intensity. This makes the differential scattering cross sections distinct from each other in the near-backward direction, but indistinguishable as we consider the near-forward scattering. Similar results have been observed considering other values of $M\omega$ and in the scalar scattering from black holes with tidal charge~\cite{deOliveira2018epjc78_876}, Reissner-Nordström~\cite{Crispino2009prd79_064022}, and Bardeen black holes~\cite{Macedo2015prd91_024012}.

In the backward direction, where the main difference of the differential scattering cross section of black holes with different tidal charges is observed, the scattering cross section is well described by the glory approximation~\cite{DeWitt-Morette1984prd29_1663,Futterman_etal1988}, which for spherically symmetric black holes can be expressed as:
\begin{equation}
 \frac{d\sigma_\text{el}^{(\text{gl})}}{d\Omega} = 2\pi \omega b_g^2 \left| \frac{db}{d\theta} \right|_{\theta = \pi} J_{2s}^2(b_g\, \omega \sin\theta),
\label{gl_scs}
\end{equation}
where $J_{n}$ is a Bessel function of the first kind~\cite{Gradshteyn_etal2000}, $s$ is the particle's spin, and $b_g$ is the impact parameter of rays scattered at $\theta = 180^\circ$.

In order to complete the scattering description via the glory approximation, we need to determine the parameters $b_g$ and $|db/d\theta|_{\theta = \pi}$. This demands a geodesic analysis, which has been made in detail in Ref.~\cite{deOliveira2018epjc78_876}. There, we showed that the increase of the tidal-charge intensity, i.e. $-q$, results in a increase of $b_g$ (see Fig. 1 of Ref.~\cite{deOliveira2018epjc78_876}). Therefore, we expect the interference-fringe widths of the differential scattering cross sections to decrease with the increase of $-q$. Furthermore, the product $b_g^2|db/d\theta|_{\theta = \pi}$, which is related to the glory intensity, also increases with the increase of $-q$.

\begin{figure}[!htbp]
 \centering
 \includegraphics[width=0.49\textwidth]{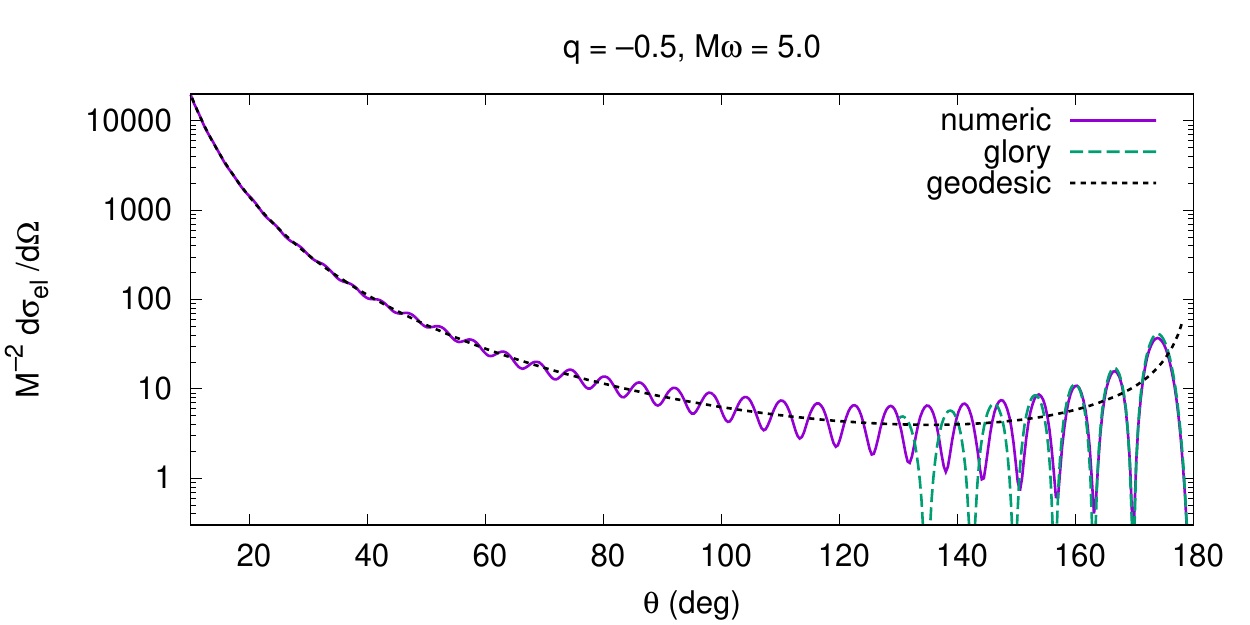}
 \includegraphics[width=0.49\textwidth]{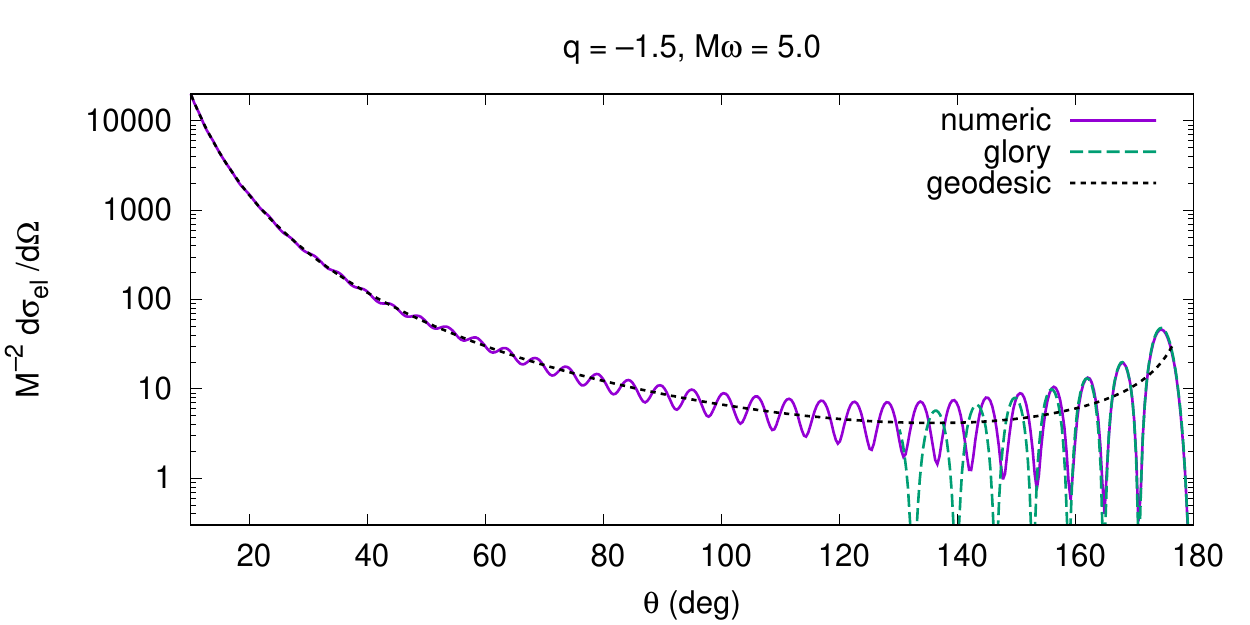}
 \caption{Comparison of the numeric results for the scattering cross sections with the glory approximation and geodesic differential scattering cross section. We consider black holes with tidal charges $q = - 0.5$ (left) and $q = -1.5$ (right).}
 \label{fig:comp_approx}
\end{figure}

We compare the numeric results for the differential scattering cross sections with the glory approximation and geodesic cross section in Fig.~\ref{fig:comp_approx}. Although both geodesic and glory results are high-frequency approximations and the chosen frequency value is not very high, these approximations agree well with the partial-wave results in their regime of validity. The glory approximation is meant to describe the interference fringes near the backward direction, so it works nicely for $\theta \gtrsim 160^\circ$. In the other hand, geodesic scattering cross section considers the deflection of scattered particles by the black hole not concerning about their interference with each other. Therefore, it works as an average value of the partial-wave result, as expected.

\begin{figure}[!hptb]
 \centering
 \includegraphics[width=0.49\textwidth]{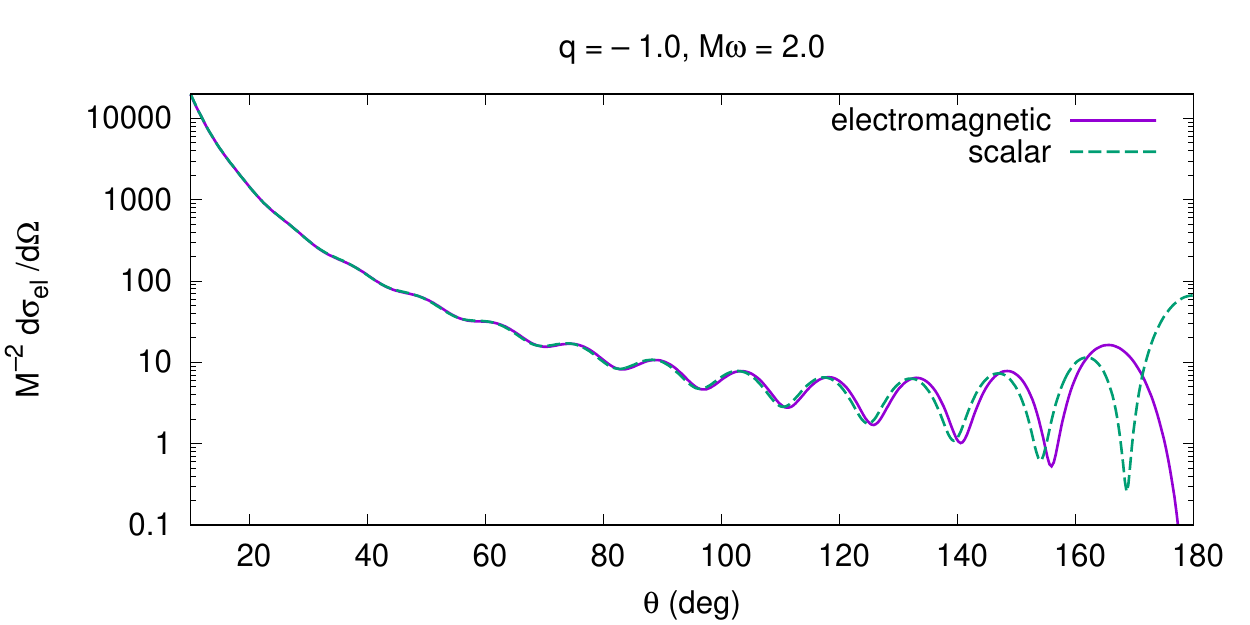}
 \includegraphics[width=0.49\textwidth]{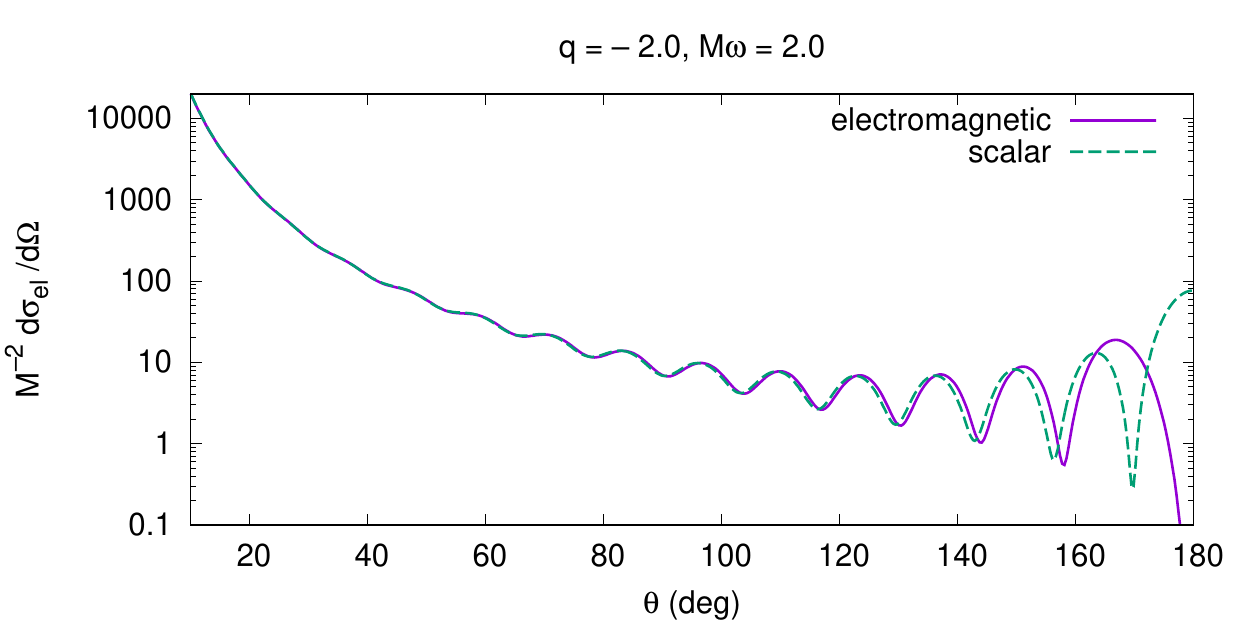}
 \caption{Comparison of the electromagnetic and scalar differential scattering cross sections considering black holes with tidal charges $q = -1.0$ (left) and $q = -2.0$ (right). The main difference occurs near the backward direction, where the flux of scattered waves is characterized by a interference maximum in the scalar case and complete destructive interference in the electromagnetic case.}
 \label{fig:comp_scs}
\end{figure}

In Fig.~\ref{fig:comp_scs} we compare the electromagnetic and scalar differential scattering cross sections of black holes with tidal charges $q = -2.0, -1.0$ and $M\omega = 2.0$. The scattering cross sections tend to fit each other for middle values of the scattering angle, $\theta \lesssim 60^\circ$. The results become very different, however, in the near-backward direction. In this direction, there is maximum flux of scattered scalar waves in opposite to zero-flux in the electromagnetic case, as it has been anticipated by the glory approximation. Such behavior has been observed in the case of Schwarzschild black holes in Ref.~\cite{Mashhoon1974prd10_1059,Crispino2009prl102321103}, where it has also been shown that the interference fringes of the differential scattering cross sections of massless bosonic fields tend to be in phase as the scattering angle decreases, while the interference fringes in the fermionic case tend to be in opposite phase to the ones of bosonic fields.

\section{Final Remarks}
\label{sec:conclusion}

We have computed numerically the electromagnetic absorption and scattering cross sections as well as the emission rate of black holes with tidal charge. The tidal charge plays important roles in the aspects of the cross sections which are related to the photon sphere and event horizon sizes. Therefore, as black holes with more intense tidal charges present bigger photon spheres, they absorb more and their scattering cross sections present narrower interference fringes. Such behaviors were successfully anticipated by the high-frequency approximations, computed via geodesic analysis and glory scattering formula which were extensively described in Ref.~\cite{deOliveira2018epjc78_876} and confirmed by the partial-wave results via comparisons made in Fig.~\ref{fig:tacs}, for the absorption cross section, and in Fig.~\ref{fig:comp_approx}, for the differential scattering cross section.

Black holes with tidal charges up to $-q = 4.6$ emit more photons than black holes with less intense tidal charges (Fig.~\ref{fig:hr}), despite the fact that the black hole temperature decreases with the increase of $-q$. This happens because black holes with more intense tidal charges are bigger, then presenting bigger absorption cross sections. Starting from $- q = 4.6$, the increase of the tidal-charge intensity results in lower total electromagnetic emission rates.

Our analysis also show that it would be hard to distinguish between black holes with tidal charge and Schwarzschild black holes via the electromagnetic scattering in the weak-field limit. This conclusion is based on the fact that the differential scattering cross section tends to depend less on the tidal-charge intensity for smaller angles, as it was shown in Fig.~\ref{fig:scs_w1}. The same conclusion has also been obtained in the analysis of null-geodesic deflection and scalar scattering~\cite{deOliveira2018epjc78_876}.

We have compared the results for both electromagnetic and scalar waves. This comparison allow us to say that, although being an idealist model, the scalar cross sections can mimic the electromagnetic results in some limits: middle-to-high frequencies in the case of absorption cross sections (Fig.~\ref{fig:abs_comp}) and middle-to-low scattering angles in the case of the differential scattering cross sections (Fig.~\ref{fig:comp_scs}).

Finally, we would like to make a parallel between the scattering cross sections presented here and their equivalents in the Reissner-Nordström case, which were presented in Ref.~\cite{Crispino2014prd90_064027}. Although black holes with tidal charge and Reissner-Nordström black holes are expressed mathematically in similar ways, these systems are drastically different. Reissner-Nordström black holes are surrounded by an electrostatic filed which is not present around the black holes analyzed here. This accounts for a qualitative difference between the cross sections presented here and the ones for Reissner-Nordström black holes, as in this last case the scattering carries the imprints of helicity reversion and conversion of electromagnetic to gravitational waves~\cite{Gerlach1974prl32_1023,Zerilli1974prd9_860,Moncrief1974prd9_2707,Moncrief1975prd12_1526,Moncrief1974prd10_1057,Olson1974prl33_1116,Matzner1976prd14_3274}. Therefore, the electromagnetic scattering from Reissner-Nordström black holes is characterized by nonzero flux in the backward direction, while for the case of black holes with tidal charge the backscattered flux is exactly zero. In visual terms, this means that the glory effect, when considering electromagnetic waves, would be seen as a bright spot in the center of concentric rings in the extreme Reissner-Nordström case~\cite{Crispino2014prd90_064027} but only bright concentric rings, no bright spot, in the case of black holes with tidal charge.

\section*{Acknowledgments}
The author would like to thank Conselho Nacional de Desenvolvimento Científico e Tecnológico (CNPq) for partial financial support via the grants 304679/2018-6 and 427532/2018-3.


\begin{thebibliography}{74}
\providecommand{\natexlab}[1]{#1}
\providecommand{\url}[1]{\texttt{#1}}
\expandafter\ifx\csname urlstyle\endcsname\relax
  \providecommand{\doi}[1]{doi: #1}\else
  \providecommand{\doi}{doi: \begingroup \urlstyle{rm}\Url}\fi

\bibitem[{\relax EHT Collaboration}(2019{\natexlab{a}})]{eht2019aj_l1}
{\relax EHT Collaboration}.
\newblock {First M87 Event Horizon Telescope Results. I. The Shadow of the
  Supermassive Black Hole}.
\newblock \emph{Astrophys. J.}, 875\penalty0 (1):\penalty0 L1,
  2019{\natexlab{a}}.
\newblock \doi{10.3847/2041-8213/ab0ec7}.

\bibitem[{\relax EHT Collaboration}(2019{\natexlab{b}})]{eht2019aj_l2}
{\relax EHT Collaboration}.
\newblock {First M87 Event Horizon Telescope Results. II. Array and
  Instrumentation}.
\newblock \emph{Astrophys. J.}, 875\penalty0 (1):\penalty0 L2,
  2019{\natexlab{b}}.
\newblock \doi{10.3847/2041-8213/ab0c96}.

\bibitem[{\relax EHT Collaboration}(2019{\natexlab{c}})]{eht2019aj_l3}
{\relax EHT Collaboration}.
\newblock {First M87 Event Horizon Telescope Results. III. Data Processing and
  Calibration}.
\newblock \emph{Astrophys. J.}, 875\penalty0 (1):\penalty0 L3,
  2019{\natexlab{c}}.
\newblock \doi{10.3847/2041-8213/ab0c57}.

\bibitem[{\relax EHT Collaboration}(2019{\natexlab{d}})]{eht2019aj_l4}
{\relax EHT Collaboration}.
\newblock {First M87 Event Horizon Telescope Results. IV. Imaging the Central
  Supermassive Black Hole}.
\newblock \emph{Astrophys. J.}, 875\penalty0 (1):\penalty0 L4,
  2019{\natexlab{d}}.
\newblock \doi{10.3847/2041-8213/ab0e85}.

\bibitem[{\relax EHT Collaboration}(2019{\natexlab{e}})]{eht2019aj_l5}
{\relax EHT Collaboration}.
\newblock {First M87 Event Horizon Telescope Results. V. Physical Origin of the
  Asymmetric Ring}.
\newblock \emph{Astrophys. J.}, 875\penalty0 (1):\penalty0 L5,
  2019{\natexlab{e}}.
\newblock \doi{10.3847/2041-8213/ab0f43}.

\bibitem[{\relax EHT Collaboration}(2019{\natexlab{f}})]{eht2019aj_l6}
{\relax EHT Collaboration}.
\newblock {First M87 Event Horizon Telescope Results. VI. The Shadow and Mass
  of the Central Black Hole}.
\newblock \emph{Astrophys. J.}, 875\penalty0 (1):\penalty0 L6,
  2019{\natexlab{f}}.
\newblock \doi{10.3847/2041-8213/ab1141}.

\bibitem[{\relax LIGO and VIRGO
  collaborations}(2016{\natexlab{a}})]{ligo2016prl116_061102}
{\relax LIGO and VIRGO collaborations}.
\newblock {Observation of Gravitational Waves from a Binary Black Hole Merger}.
\newblock \emph{Phys. Rev. Lett.}, 116\penalty0 (6):\penalty0 061102,
  2016{\natexlab{a}}.
\newblock \doi{10.1103/PhysRevLett.116.061102}.

\bibitem[{\relax LIGO and VIRGO
  collaborations}(2016{\natexlab{b}})]{ligo2016prl116_241103}
{\relax LIGO and VIRGO collaborations}.
\newblock {GW151226: Observation of Gravitational Waves from a 22-Solar-Mass
  Binary Black Hole Coalescence}.
\newblock \emph{Phys. Rev. Lett.}, 116\penalty0 (24):\penalty0 241103,
  2016{\natexlab{b}}.
\newblock \doi{10.1103/PhysRevLett.116.241103}.

\bibitem[{\relax LIGO and VIRGO collaborations}(2017)]{ligo2017prl118_221101}
{\relax LIGO and VIRGO collaborations}.
\newblock {GW170104: Observation of a 50-Solar-Mass Binary Black Hole
  Coalescence at Redshift 0.2}.
\newblock \emph{Phys. Rev. Lett.}, 118\penalty0 (22):\penalty0 221101, 2017.
\newblock \doi{10.1103/PhysRevLett.118.221101}.

\bibitem[Broderick and Loeb(2006)]{Broderick2006aj636_l109}
Avery~E. Broderick and Abraham Loeb.
\newblock {Frequency-dependent Shift in the Image Centroid of the Black Hole at
  the Galactic Center as a Test of General Relativity}.
\newblock \emph{Astrophys. J.}, 636:\penalty0 L109--L112, 2006.
\newblock \doi{10.1086/500008}.

\bibitem[Moscibrodzka et~al.(2009)Moscibrodzka, Gammie, Dolence, Shiokawa, and
  Leung]{Moscibrodzka2009aj706_497}
Monika Moscibrodzka, Charles~F. Gammie, Joshua~C. Dolence, Hotaka Shiokawa, and
  Po~Kin Leung.
\newblock {Radiative Models of Sgr A* from GRMHD Simulations}.
\newblock \emph{Astrophys. J.}, 706:\penalty0 497--507, 2009.
\newblock \doi{10.1088/0004-637X/706/1/497}.

\bibitem[Dexter et~al.(2012)Dexter, McKinney, and
  Agol]{Dexter2012mnras421_1517}
Jason Dexter, Jonathan~C. McKinney, and Eric Agol.
\newblock {The size of the jet launching region in M87}.
\newblock \emph{Mon. Not. Roy. Astron. Soc.}, 421:\penalty0 1517, 2012.
\newblock \doi{10.1111/j.1365-2966.2012.20409.x}.

\bibitem[Dibi et~al.(2012)Dibi, Drappeau, Fragile, Markoff, and
  Dexter]{Dibi2012mnras426_1928}
Salome Dibi, Samia Drappeau, P.~Chris Fragile, Sera Markoff, and Jason Dexter.
\newblock {General relativistic magnetohydrodynamic simulations of accretion on
  to Sgr A*: how important are radiative losses?}
\newblock \emph{Mon. Not. Roy. Astron. Soc.}, 426:\penalty0 1928, 2012.
\newblock \doi{10.1111/j.1365-2966.2012.21857.x}.

\bibitem[Chan et~al.(2015)Chan, Psaltis, Özel, Narayan, and
  Sądowski]{Chan2015aj799_1}
Chi-Kwan Chan, Dimitrios Psaltis, Feryal Özel, Ramesh Narayan, and Aleksander
  Sądowski.
\newblock {The Power of Imaging: Constraining the Plasma Properties of GRMHD
  Simulations using EHT Observations of Sgr A*}.
\newblock \emph{Astrophys. J.}, 799\penalty0 (1):\penalty0 1, 2015.
\newblock \doi{10.1088/0004-637X/799/1/1}.

\bibitem[Moscibrodzka et~al.(2016)Moscibrodzka, Falcke, and
  Shiokawa]{Moscibrodzka2016aa586_a38}
Monika Moscibrodzka, Heino Falcke, and Hotaka Shiokawa.
\newblock {General relativistic magnetohydrodynamical simulations of the jet in
  M 87}.
\newblock \emph{Astron. Astrophys.}, 586:\penalty0 A38, 2016.
\newblock \doi{10.1051/0004-6361/201526630}.

\bibitem[Porth et~al.(2016)Porth, Olivares, Mizuno, Younsi, Rezzolla,
  Moscibrodzka, Falcke, and Kramer]{Porth2016cac4_1}
Oliver Porth, Hector Olivares, Yosuke Mizuno, Ziri Younsi, Luciano Rezzolla,
  Monika Moscibrodzka, Heino Falcke, and Michael Kramer.
\newblock {The black hole accretion code}.
\newblock \emph{Compu. Astrophys. and Cosmology}, 4:\penalty0 1, 2016.
\newblock \doi{10.1186/s40668-017-0020-2}.

\bibitem[Chael et~al.(2018)Chael, Narayan, and Johnson]{Chael2018mnras}
Andrew Chael, Ramesh Narayan, and Michael~D. Johnson.
\newblock {Two-temperature, Magnetically Arrested Disc simulations of the jet
  from the supermassive black hole in M87}.
\newblock 2018.
\newblock \doi{10.1093/mnras/stz988}.

\bibitem[Ryan et~al.(2018)Ryan, Ressler, Dolence, Gammie, and
  Quataert]{Ryan2018aj864_126}
Benjamin~R. Ryan, Sean~M. Ressler, Joshua~C. Dolence, Charles~F. Gammie, and
  Eliot Quataert.
\newblock {Two-temperature GRRMHD Simulations of M87}.
\newblock \emph{Astrophys. J.}, 864\penalty0 (2):\penalty0 126, 2018.
\newblock \doi{10.3847/1538-4357/aad73a}.

\bibitem[Buonanno and Damour(2000)]{Buonanno2000prd62_064015}
Alessandra Buonanno and Thibault Damour.
\newblock {Transition from inspiral to plunge in binary black hole
  coalescences}.
\newblock \emph{Phys. Rev. D}, 62:\penalty0 064015, 2000.
\newblock \doi{10.1103/PhysRevD.62.064015}.

\bibitem[Centrella et~al.(2010)Centrella, Baker, Kelly, and van
  Meter]{Centrella2010rmp82_3069}
Joan Centrella, John~G. Baker, Bernard~J. Kelly, and James~R. van Meter.
\newblock {Black-hole binaries, gravitational waves, and numerical relativity}.
\newblock \emph{Rev. Mod. Phys.}, 82:\penalty0 3069, 2010.
\newblock \doi{10.1103/RevModPhys.82.3069}.

\bibitem[Taracchini et~al.(2012)Taracchini, Pan, Buonanno, Barausse, Boyle,
  Chu, Lovelace, Pfeiffer, and Scheel]{Taracchini2012prd86_024011}
Andrea Taracchini, Yi~Pan, Alessandra Buonanno, Enrico Barausse, Michael Boyle,
  Tony Chu, Geoffrey Lovelace, Harald~P. Pfeiffer, and Mark~A. Scheel.
\newblock {Prototype effective-one-body model for nonprecessing spinning
  inspiral-merger-ringdown waveforms}.
\newblock \emph{Phys. Rev. D}, 86:\penalty0 024011, 2012.
\newblock \doi{10.1103/PhysRevD.86.024011}.

\bibitem[Pan et~al.(2011)Pan, Buonanno, Boyle, Buchman, Kidder, Pfeiffer, and
  Scheel]{Pan2011prd84_124052}
Yi~Pan, Alessandra Buonanno, Michael Boyle, Luisa~T. Buchman, Lawrence~E.
  Kidder, Harald~P. Pfeiffer, and Mark~A. Scheel.
\newblock {Inspiral-merger-ringdown multipolar waveforms of nonspinning
  black-hole binaries using the effective-one-body formalism}.
\newblock \emph{Phys. Rev. D}, 84:\penalty0 124052, 2011.
\newblock \doi{10.1103/PhysRevD.84.124052}.

\bibitem[Littenberg et~al.(2013)Littenberg, Baker, Buonanno, and
  Kelly]{Littenberg2013prd87_104003}
Tyson~B. Littenberg, John~G. Baker, Alessandra Buonanno, and Bernard~J. Kelly.
\newblock {Systematic biases in parameter estimation of binary black-hole
  mergers}.
\newblock \emph{Phys. Rev. D}, 87\penalty0 (10):\penalty0 104003, 2013.
\newblock \doi{10.1103/PhysRevD.87.104003}.

\bibitem[Taracchini et~al.(2014)]{Taracchini2014prd89_061502}
Andrea Taracchini et~al.
\newblock {Effective-one-body model for black-hole binaries with generic mass
  ratios and spins}.
\newblock \emph{Phys. Rev. D}, 89\penalty0 (6):\penalty0 061502, 2014.
\newblock \doi{10.1103/PhysRevD.89.061502}.

\bibitem[Pürrer(2014)]{Purrer2014cqg31_195010}
Michael Pürrer.
\newblock {Frequency-domain reduced order models for gravitational waves from
  aligned-spin compact binaries}.
\newblock \emph{Class. Quant. Grav.}, 31\penalty0 (19):\penalty0 195010, 2014.
\newblock \doi{10.1088/0264-9381/31/19/195010}.

\bibitem[Dadhich et~al.(2000)Dadhich, Maartens, Papadopoulos, and
  Rezania]{Dadhich2000plb487_1}
Naresh Dadhich, Roy Maartens, Philippos Papadopoulos, and Vahid Rezania.
\newblock {Black holes on the brane}.
\newblock \emph{Phys. Lett. B}, 487:\penalty0 1--6, 2000.
\newblock \doi{10.1016/S0370-2693(00)00798-X}.

\bibitem[Randall and Sundrum(1999{\natexlab{a}})]{Randall1999prl83_3370}
Lisa Randall and Raman Sundrum.
\newblock {Large Mass Hierarchy from a Small Extra Dimension}.
\newblock \emph{Phys. Rev. Lett.}, 83:\penalty0 3370--3373, 1999{\natexlab{a}}.
\newblock \doi{10.1103/PhysRevLett.83.3370}.

\bibitem[Randall and Sundrum(1999{\natexlab{b}})]{Randall1999prl83_4690}
Lisa Randall and Raman Sundrum.
\newblock {An Alternative to Compactification}.
\newblock \emph{Phys. Rev. Lett.}, 83:\penalty0 4690--4693, 1999{\natexlab{b}}.
\newblock \doi{10.1103/PhysRevLett.83.4690}.

\bibitem[Chandrasekhar(1983)]{Chandra1983}
Subrahmanyan Chandrasekhar.
\newblock \emph{The Mathematical Theory of Black Holes}.
\newblock Clarendon Press, Oxford, 1983.
\newblock ISBN 0-19-851291-0.

\bibitem[{\relax ATLAS Collaboration}(2016{\natexlab{a}})]{ATLAS2016jhep03_041}
{\relax ATLAS Collaboration}.
\newblock {Search for new phenomena with photon+jet events in proton-proton
  collisions at $ \sqrt{s}=13 $ TeV with the ATLAS detector}.
\newblock \emph{JHEP}, 03:\penalty0 041, 2016{\natexlab{a}}.
\newblock \doi{10.1007/JHEP03(2016)041}.

\bibitem[{\relax ATLAS Collaboration}(2016{\natexlab{b}})]{ATLAS2016plb754_302}
{\relax ATLAS Collaboration}.
\newblock {Search for new phenomena in dijet mass and angular distributions
  from $pp$ collisions at $\sqrt{s}=$ 13 TeV with the ATLAS detector}.
\newblock \emph{Phys. Lett. B}, 754:\penalty0 302--322, 2016{\natexlab{b}}.
\newblock \doi{10.1016/j.physletb.2016.01.032}.

\bibitem[{\relax CMS Collaboration}(2017)]{CMS2017plb774_279}
{\relax CMS Collaboration}.
\newblock {Search for black holes and other new phenomena in high-multiplicity
  final states in proton–proton collisions at $ \sqrt{s}=$13 TeV}.
\newblock \emph{Phys. Lett. B}, 774:\penalty0 279--307, 2017.
\newblock \doi{10.1016/j.physletb.2017.09.053}.

\bibitem[{\relax ATLAS Collaboration}(2018)]{ATLAS2018epjc78_102}
{\relax ATLAS Collaboration}.
\newblock {Search for new phenomena in high-mass final states with a photon and
  a jet from $pp$ collisions at $\sqrt{s}$ = 13 TeV with the ATLAS detector}.
\newblock \emph{Eur. Phys. J. C}, 78\penalty0 (2):\penalty0 102, 2018.
\newblock \doi{10.1140/epjc/s10052-018-5553-2}.

\bibitem[{\relax CMS Collaboration}(2018)]{cms2018jhep2018_42}
{\relax CMS Collaboration}.
\newblock {Search for black holes and sphalerons in high-multiplicity final
  states in proton-proton collisions at $ \sqrt{s}=13 $ TeV}.
\newblock \emph{JHEP}, 2018\penalty0 (11):\penalty0 042, 2018.
\newblock \doi{10.1007/JHEP11(2018)042}.

\bibitem[Arkani-Hamed et~al.(1998)Arkani-Hamed, Dimopoulos, and
  Dvali]{Arkani1998plb429_263}
Nima Arkani-Hamed, Savas Dimopoulos, and Gia Dvali.
\newblock {The hierarchy problem and new dimensions at a millimeter}.
\newblock \emph{Phys. Lett. B}, 429:\penalty0 263--272, 1998.
\newblock \doi{10.1016/S0370-2693(98)00466-3}.

\bibitem[Antoniadis et~al.(1998)Antoniadis, Arkani-Hamed, Dimopoulos, and
  Dvali]{Antoniadis1998plb436_257}
Ignatios Antoniadis, Nima Arkani-Hamed, Savas Dimopoulos, and Gia Dvali.
\newblock {New dimensions at a millimeter to a fermi and superstrings at a
  TeV}.
\newblock \emph{Phys. Lett. B}, 436:\penalty0 257--263, 1998.
\newblock \doi{10.1016/S0370-2693(98)00860-0}.

\bibitem[de~Oliveira(2018)]{deOliveira2018epjc78_876}
Ednilton~S. de~Oliveira.
\newblock {Scalar scattering from black holes with tidal charge}.
\newblock \emph{Eur. Phys. J. C}, 78\penalty0 (11):\penalty0 876, 2018.
\newblock \doi{10.1140/epjc/s10052-018-6316-9}.

\bibitem[Sanchez(1976)]{Sanchez1976jmp17_688}
Norma~G. Sanchez.
\newblock {Scattering of scalar waves from a Schwarzschild black hole}.
\newblock \emph{J. Math. Phys.}, 17\penalty0 (5):\penalty0 688, 1976.
\newblock \doi{10.1063/1.522949}.

\bibitem[Sánchez(1977)]{Sanchez1976prd16_937}
Norma Sánchez.
\newblock {Wave scattering theory and the absorption problem for a black hole}.
\newblock \emph{Phys. Rev. D}, 16:\penalty0 937--945, 1977.
\newblock \doi{10.1103/PhysRevD.16.937}.

\bibitem[Sanchez(1978)]{Sanchez1978prd18_1030}
Norma Sanchez.
\newblock {Absorption and emission spectra of a Schwarzschild black hole}.
\newblock \emph{Phys. Rev. D}, 18:\penalty0 1030, 1978.
\newblock \doi{10.1103/PhysRevD.18.1030}.

\bibitem[Sánchez(1978)]{Sanchez1978prd18_1798}
Norma Sánchez.
\newblock {Elastic scattering of waves by a black hole}.
\newblock \emph{Phys. Rev. D}, 18:\penalty0 1798, 1978.
\newblock \doi{10.1103/PhysRevD.18.1798}.

\bibitem[Jung and Park(2005)]{Jung2005npb717_272}
Eylee Jung and D.~K. Park.
\newblock {Absorption and emission spectra of an higher-dimensional
  Reissner-Nordström black hole}.
\newblock \emph{Nucl. Phys. B}, 717:\penalty0 272--303, 2005.
\newblock \doi{10.1016/j.nuclphysb.2005.03.037}.

\bibitem[Crispino et~al.(2009{\natexlab{a}})Crispino, Dolan, and
  Oliveira]{Crispino2009prd79_064022}
Luís C.~B. Crispino, Sam~R. Dolan, and Ednilton~S. Oliveira.
\newblock {Scattering of massless scalar waves by Reissner-Nordstr\"om black
  holes}.
\newblock \emph{Phys. Rev. D}, 79:\penalty0 064022, 2009{\natexlab{a}}.
\newblock \doi{10.1103/PhysRevD.79.064022}.

\bibitem[Schee and Stuchlík(2009)]{Schee2008ijmpd18_983}
Jan Schee and Zdeněk Stuchlík.
\newblock Optical phenomena in the field of braneworld kerr black holes.
\newblock \emph{Int. J. Mod. Phys. D}, 18:\penalty0 983--1024, 2009.
\newblock \doi{10.1142/S0218271809014881}.

\bibitem[Amarilla and Eiroa(2012)]{Amarilla2011pprd85_064019}
Leonardo Amarilla and Ernesto~F. Eiroa.
\newblock {Shadow of a rotating braneworld black hole}.
\newblock \emph{Phys. Rev. D}, 85:\penalty0 064019, 2012.
\newblock \doi{10.1103/PhysRevD.85.064019}.

\bibitem[Abdujabbarov et~al.(2017)Abdujabbarov, Ahmedov, Dadhich, and
  Atamurotov]{Abdujabbarov2017prd96_084017}
Ahmadjon Abdujabbarov, Bobomurat Ahmedov, Naresh Dadhich, and Farruh
  Atamurotov.
\newblock {Optical properties of a braneworld black hole: Gravitational lensing
  and retrolensing}.
\newblock \emph{Phys. Rev. D}, 96\penalty0 (8):\penalty0 084017, 2017.
\newblock \doi{10.1103/PhysRevD.96.084017}.

\bibitem[Eiroa and Sendra(2018)]{Eiroa2017epjc78_91}
Ernesto~F. Eiroa and Carlos~M. Sendra.
\newblock {Shadow cast by rotating braneworld black holes with a cosmological
  constant}.
\newblock \emph{Eur. Phys. J. C}, 78\penalty0 (2):\penalty0 91, 2018.
\newblock \doi{10.1140/epjc/s10052-018-5586-6}.

\bibitem[Toshmatov et~al.(2016)Toshmatov, Stuchlík, Schee, and
  Ahmedov]{Toshmatov2016prd93_124017}
Bobir Toshmatov, Zdeněk Stuchlík, Jan Schee, and Bobomurat Ahmedov.
\newblock {Quasinormal frequencies of black hole in the braneworld}.
\newblock \emph{Phys. Rev. D}, 93\penalty0 (12):\penalty0 124017, 2016.
\newblock \doi{10.1103/PhysRevD.93.124017}.

\bibitem[Abdujabbarov and Ahmedov(2010)]{Abdujabbarov2010prd81_044022}
Ahmadjon Abdujabbarov and Bobomurat Ahmedov.
\newblock {Test particle motion around a black hole in a braneworld}.
\newblock \emph{Phys. Rev. D}, 81:\penalty0 044022, 2010.
\newblock \doi{10.1103/PhysRevD.81.044022}.

\bibitem[Rahimov et~al.(2011)Rahimov, Abdujabbarov, and
  Ahmedov]{Rahimov2011ass335_499}
O.~G. Rahimov, A.~A. Abdujabbarov, and B.~J. Ahmedov.
\newblock {Magnetized particle capture cross section for braneworld black
  hole}.
\newblock \emph{Astrophys. Space Sci.}, 335:\penalty0 499--504, 2011.
\newblock \doi{10.1007/s10509-011-0755-1}.

\bibitem[Shaymatov et~al.(2013)Shaymatov, Ahmedov, and
  Abdujabbarov]{Shaymatov2013prd88_024016}
S.~R. Shaymatov, B.~J. Ahmedov, and A.~A. Abdujabbarov.
\newblock {Particle acceleration near a rotating black hole in a
  Randall-Sundrum brane with a cosmological constant}.
\newblock \emph{Phys. Rev. D}, 88\penalty0 (2):\penalty0 024016, 2013.
\newblock \doi{10.1103/PhysRevD.88.024016}.

\bibitem[Crispino et~al.(2007)Crispino, Oliveira, Higuchi, and
  Matsas]{Crispino2007prd75_104012}
Luis C.~B. Crispino, Ednilton~S. Oliveira, Atsushi Higuchi, and George E.~A.
  Matsas.
\newblock {Absorption cross section of electromagnetic waves for Schwarzschild
  black holes}.
\newblock \emph{Phys. Rev. D}, 75:\penalty0 104012, 2007.
\newblock \doi{10.1103/PhysRevD.75.104012}.

\bibitem[Higuchi(1987)]{Higuchi1987cqg4_721}
Atsushi Higuchi.
\newblock {Quantization of Scalar and Vector Fields Inside the Cosmological
  Event Horizon and Its Application to Hawking Effect}.
\newblock \emph{Class. Quant. Grav.}, 4:\penalty0 721, 1987.
\newblock \doi{10.1088/0264-9381/4/3/029}.

\bibitem[Molina et~al.(2016)Molina, Pavan, and
  Medina~Torrejón]{Molina2016prd93_124068}
C.~Molina, A.~B. Pavan, and T.~E. Medina~Torrejón.
\newblock {Electromagnetic perturbations in new brane world scenarios}.
\newblock \emph{Phys. Rev. D}, 93\penalty0 (12):\penalty0 124068, 2016.
\newblock \doi{10.1103/PhysRevD.93.124068}.

\bibitem[Crispino et~al.(2009{\natexlab{b}})Crispino, Dolan, and
  Oliveira]{Crispino2009prl102321103}
Luis C.~B. Crispino, Sam~R. Dolan, and Ednilton~S. Oliveira.
\newblock {Electromagnetic wave scattering by Schwarzschild black holes}.
\newblock \emph{Phys. Rev. Lett.}, 102:\penalty0 231103, 2009{\natexlab{b}}.
\newblock \doi{10.1103/PhysRevLett.102.231103}.

\bibitem[Crispino et~al.(2009{\natexlab{c}})Crispino, Higuchi, and
  Oliveira]{Crispino2009prd80_104026}
Luis C.~B. Crispino, Atsushi Higuchi, and Ednilton~S. Oliveira.
\newblock {Electromagnetic absorption cross section of Reissner-Nordstr\"om
  black holes revisited}.
\newblock \emph{Phys. Rev. D}, 80:\penalty0 104026, 2009{\natexlab{c}}.
\newblock \doi{10.1103/PhysRevD.80.104026}.

\bibitem[Hawking(1975)]{Hawking1975cmp43_199}
S.~W. Hawking.
\newblock {Particle creation by black holes}.
\newblock \emph{Commun. Math. Phys.}, 43:\penalty0 199--220, 1975.
\newblock \doi{10.1007/BF02345020}.
\newblock [,167(1975)].

\bibitem[Fabbri(1975)]{Fabbri1975prd12_933}
R.~Fabbri.
\newblock {Scattering and absorption of electromagnetic waves by a
  Schwarzschild black hole}.
\newblock \emph{Phys. Rev. D}, 12:\penalty0 933--942, 1975.
\newblock \doi{10.1103/PhysRevD.12.933}.

\bibitem[Crispino et~al.(2014)Crispino, Dolan, Higuchi, and
  de~Oliveira]{Crispino2014prd90_064027}
Luís C.~B. Crispino, Sam~R. Dolan, Atsushi Higuchi, and Ednilton~S.
  de~Oliveira.
\newblock {Inferring black hole charge from backscattered electromagnetic
  radiation}.
\newblock \emph{Phys. Rev. D}, 90\penalty0 (6):\penalty0 064027, 2014.
\newblock \doi{10.1103/PhysRevD.90.064027}.

\bibitem[Mashhoon(1973)]{Mashhoon1973prd7_2807}
Bahram Mashhoon.
\newblock {Scattering of Electromagnetic Radiation from a Black Hole}.
\newblock \emph{Phys. Rev. F}, 7:\penalty0 2807--2814, 1973.
\newblock \doi{10.1103/PhysRevD.7.2807}.

\bibitem[Abramowitz and Stegun(1965)]{Abramowitz_etal1964}
M.~Abramowitz and I.~A. Stegun.
\newblock \emph{Handbook of Mathematical Functions}.
\newblock Dover Publications, New York, 1965.

\bibitem[Yennie et~al.(1954)Yennie, Ravenhall, and Wilson]{Yennie1954pr85_500}
D.~R. Yennie, D.~G. Ravenhall, and R.~N. Wilson.
\newblock {Phase-Shift Calculation of High-Energy Electron Scattering}.
\newblock \emph{Phys. Rev.}, 95:\penalty0 500--512, 1954.
\newblock \doi{10.1103/PhysRev.95.500}.

\bibitem[Macedo et~al.(2015)Macedo, de~Oliveira, and
  Crispino]{Macedo2015prd91_024012}
Caio F.~B. Macedo, Ednilton~S. de~Oliveira, and Luís C.~B. Crispino.
\newblock {Scattering by regular black holes: Planar massless scalar waves
  impinging upon a Bardeen black hole}.
\newblock \emph{Phys. Rev. D}, 92\penalty0 (2):\penalty0 024012, 2015.
\newblock \doi{10.1103/PhysRevD.92.024012}.

\bibitem[DeWitt-Morette and Nelson(1984)]{DeWitt-Morette1984prd29_1663}
Cécile DeWitt-Morette and Bruce~L. Nelson.
\newblock {Glories---and other degenerate points of the action}.
\newblock \emph{Phys. Rev. D}, 29:\penalty0 1663--1668, 1984.
\newblock \doi{10.1103/PhysRevD.29.1663}.

\bibitem[Futterman et~al.(1988)Futterman, Handler, and
  Matzner]{Futterman_etal1988}
J.~A.~H. Futterman, F.~A. Handler, and R.~A. Matzner.
\newblock \emph{Scattering from Black Holes}.
\newblock Cambridge University Press, Cambridge, 1988.
\newblock ISBN 0-521-32986-8.

\bibitem[Gradshteyn and Ryzhik(2015)]{Gradshteyn_etal2000}
I.~S. Gradshteyn and I.~M. Ryzhik.
\newblock \emph{Table of Integrals, Series, and Products}.
\newblock Academic Press, San Diego, 8 th edition, 2015.
\newblock ISBN 978-0-12-384933-5.

\bibitem[Mashhoon(1974)]{Mashhoon1974prd10_1059}
Bahram Mashhoon.
\newblock {Electromagnetic scattering from a black hole and the glory effect}.
\newblock \emph{Phys. Rev. D}, 10:\penalty0 1059--1063, 1974.
\newblock \doi{10.1103/PhysRevD.10.1059}.

\bibitem[Gerlach(1974)]{Gerlach1974prl32_1023}
Ulrich~H. Gerlach.
\newblock {Beat Frequency Oscillations near Charged Black Holes and Other
  Electrovacuum Geometries}.
\newblock \emph{Phys. Rev. Lett.}, 32:\penalty0 1023--1025, 1974.
\newblock \doi{10.1103/PhysRevLett.32.1023}.

\bibitem[Zerilli(1974)]{Zerilli1974prd9_860}
F.~J. Zerilli.
\newblock {Perturbation analysis for gravitational and electromagnetic
  radiation in a Reissner-Nordstr\"om geometry}.
\newblock \emph{Phys. Rev. D}, 9:\penalty0 860--868, 1974.
\newblock \doi{10.1103/PhysRevD.9.860}.

\bibitem[Moncrief(1974{\natexlab{a}})]{Moncrief1974prd9_2707}
Vincent Moncrief.
\newblock {Odd-parity stability of a Reissner-Nordstr\"om black hole}.
\newblock \emph{Phys. Rev. D}, 9:\penalty0 2707--2709, 1974{\natexlab{a}}.
\newblock \doi{10.1103/PhysRevD.9.2707}.

\bibitem[Moncrief(1975)]{Moncrief1975prd12_1526}
Vincent Moncrief.
\newblock {Gauge-invariant perturbations of Reissner-Nordstr\"om black holes}.
\newblock \emph{Phys. Rev. D}, 12:\penalty0 1526--1537, 1975.
\newblock \doi{10.1103/PhysRevD.12.1526}.

\bibitem[Moncrief(1974{\natexlab{b}})]{Moncrief1974prd10_1057}
Vincent Moncrief.
\newblock {Stability of Reissner-Nordstr\"om black holes}.
\newblock \emph{Phys. Rev. D}, 10:\penalty0 1057--1059, 1974{\natexlab{b}}.
\newblock \doi{10.1103/PhysRevD.10.1057}.

\bibitem[Olson and Unruh(1974)]{Olson1974prl33_1116}
D.~W. Olson and W.~G. Unruh.
\newblock {Conversion of electromagnetic to gravitational radiation by
  scattering from a charged black hole}.
\newblock \emph{Phys. Rev. Lett.}, 33:\penalty0 1116--1119, 1974.
\newblock \doi{10.1103/PhysRevLett.33.1116}.

\bibitem[Matzner(1976)]{Matzner1976prd14_3274}
R.~A. Matzner.
\newblock {Low Frequency Limit Conversion Cross-Sections for Charged Black
  Holes}.
\newblock \emph{Phys. Rev. D}, 14:\penalty0 3274--3280, 1976.
\newblock \doi{10.1103/PhysRevD.14.3274}.

\end{thebibliography}
\end{document}